\documentclass[aps,twocolumn,longbibliography,pra,superscriptaddress,amssymb,amsmath,amsmath,floatfix]{revtex4-2}
\usepackage{graphicx}
\usepackage{epstopdf}
\usepackage{comment}
\usepackage{bm}
\usepackage{wrapfig}
\usepackage{times}
\usepackage[colorlinks=true,linkcolor=blue,urlcolor=blue,citecolor=blue]{hyperref}
\usepackage{float}
\usepackage{color,soul}
\newenvironment{dataavailability}
  {\section*{Data Availability}}
  {}

\begin{document}
\title{The He$_2(a~^3\Sigma_u^+)$+He complex: A promising system for cold and ultracold collision studies
}

\author{Dibyendu Sardar}
\email{dibyendu.sardar@univ-rennes.fr}
\affiliation{Institut de Physique de Rennes, CNRS-Universit\'e de Rennes, UMR 6251, F-35000 Rennes, France}

\author{Fran\c{c}ois Lique}
\email{francois.lique@univ-rennes.fr}
\affiliation{Institut de Physique de Rennes, CNRS-Universit\'e de Rennes, UMR 6251, F-35000 Rennes, France}

\date{\today}% It is always \today, today,
\begin{abstract}
Metastable helium dimer, He$_2$($a~^3\Sigma_u^+$), is a promising molecular system for direct laser cooling and precision spectroscopy. Interaction of He$_2$($a~^3\Sigma_u^+$) with a ground-state He to form a bound trimer remains an intriguing and unexplored question in few-body quantum physics. Here, we investigate intermolecular interactions, and explore spectroscopic and quantum scattering properties of He$_2(a~^3\Sigma_u^+)$+He complex. We calculate potential interactions for the He$_3$ trimer using state-of-the-art \textit{ab initio} methods of quantum chemistry. We employ the gold standard coupled cluster approach together with large atomic basis sets. We predict a linear equilibrium geometrical shape of He$_3$ characterized by an exceptionally shallow potential well of approximately 1.084 cm$^{-1}$. Quantum close-coupling calculations, explicitly accounting for the fine-structure of He$_2$($a~^3\Sigma_u^+$), predict a single bound state with a binding energy of 9.423 mK. This bound state persists under substantial variation of the interaction potential. Energy dependent inelastic scattering cross sections exhibit smooth variation with collision energy associated with the sparse bound-state spectrum of the trimer. Our results establish a theoretical benchmark for unique quantum properties of metastable He$_3$. Predicted molecular properties of metastable He$_3$ are promising for future experimental investigations, particularly through advanced Coulomb-explosion imaging techniques.
\end{abstract}

%We investigate the quantum properties of the metastable helium trimer, He$_3$. He$_3$ is obtanied by the interaction of a spin-polarized metastable helium dimer, He$_2$($a~^3\Sigma_u^+$), with a ground-state helium atom, He($^1\mathrm{S}$). A highly accurate potential energy surface is developed using state-of-the-art coupled-cluster and multireference configuration-interaction methods with extensive Gaussian basis sets. Independent electronic-structure calculations consistently predict a linear equilibrium configuration, characterized by an exceptionally shallow potential well of approximately 1.05 cm$^{-1}$. Quantum close-coupling calculations, explicitly accounting for the fine structure of the metastable helium dimer, reveal the existence of a single bound state with a binding energy of 8.337 mK. Remarkably, this bound state persists under substantial variations of the interaction potential, highlighting the robustness of the prediction against uncertainties in the underlying electronic-structure calculations. Our results establish a theoretical benchmark for a metastable helium trimer that supports only one bound quantum state and provide strong motivation for its experimental observation. In particular, the predicted binding energy and linear molecular structure offer promising targets for future investigations using advanced Coulomb-explosion imaging techniques.

\maketitle
%Bound-state calculations reveal six bound levels of the He$_2$+Ne complex, from which we determine its spectroscopic properties.
%Calculated bound-state energies, spectroscopic constants, and scattering resonance characteristics provide a quantitative description of the He$_2$+Ne van der Waals complex and establish a theoretical basis for the interpretation of future experimental spectra and collision measurements.

%while  multireference configuration interaction method with single and double excitations is used for He$_2(^3\Sigma^+_u)$+Ar.while we construct a 2D surface is computed for He$_2(^3\Sigma^+_u)$+Ar under rigid rotor approximation for He$_2$.

%\begin{multicols}{2}
%----------------------------------------------------------------------------------------------
%\section{Introduction}
%\label{sec:intro}
%---------------------------------------------------------------------------------------------
%\textit{Introduction}\textemdash
Helium has attracted considerable attention in the field of atomic, molecular, and optical Physics. Helium and weakly bound small helium-cluster exhibit fascinating quantum behavior unseen in other elements. The smallest aggregates of He, dimer (He$_2$) and trimer (He$_3$) are predicted to exhibit unique structural properties with extremely large amplitude motion. These systems represent natural quantum objects for investigating fundamental quantum phenomena. 

%The feasibility of first-principles quantum-chemical calculations for these light, few-electron helium clusters enables it ideal for benchmarking theoretical predictions of elastic and inelastic scattering calculations. 

The helium dimer in its molecular ground electronic state, $X~^1\Sigma_g^+$, has been observed experimentally \cite{LuoJCP1992}. Two electrons of helium atom are strongly bound, making He the most unpolarizable element. Thus, the van der Waals force between two He atoms is extremely weak. The potential well depth for ground $X~^1\Sigma_g^+$ electronic state of He$_2$ is 7.64 cm$^{-1}$ 
\cite{przybytekPRL2010}, and has a single bound state \cite{LuoJCP1992} with a binding energy of $1.76 \pm 0.15$ mK \cite{zeller2016}. The reported $s$-wave scattering length for He$_2$ dimer is $90.4$ angstrom \cite{CencekJCP2012}.

The helium cluster composed of three atoms in their ground state is an even more intriguing quantum system. First-principles quantum-mechanical calculations  predict two bound states for helium trimer \cite{HiyamaPRA2012}. The extremely weakly bound excited state is of Efimov nature verified experimentally using the Coulomb explosion imaging technique \cite{KunitskiScience2015}. Key feature of the Efimov state is its unusual extent and structure, and have importance relevance in many areas of physics. First experimental signatures of Efimov trimers have been reported for a trapped cold atom system \cite{KraemerNature2006}. Theoretical studies predicted both equilateral triangular structure and linear configuration of He$_3$, however, recent experimental advances have demonstrated that helium trimer is  a structureless random cloud \cite{VoigtsbergerNatCommn2014}. 

%using most accurate He$_2$($X~^1\Sigma_g^+$) molecular potential \cite{CencekJCP2012}
%, the ground and excited states, with binding energies $131.84$ mK and $2.6505$ mK , respectively. 

%Efimov trimer is approximately 100 times larger compared to a chemically bound triatomic molecule.

Bound state properties and structural characteristics of trimer involving metastable helium atoms remain largely unexplored, presenting an intriguing opportunity for theoretical and experimental investigations. In this context, helium dimer in its lowest-lying triplet electronic state, $a~^3\Sigma_u^+$, which correlates to the He(1~$^1\mathrm{S})+\mathrm{He}(2~^3\mathrm{S})$ dissociation limit is a novel system. The $a~^3\Sigma_u^+$ state is metastable with a lifetime of several seconds \cite{ChabalowskiJCP1989}. He$_2$($a~^3\Sigma_u^+$) is then a promising candidate to become the first homonuclear molecule to be directly laser-cooled, due to its near-diagonal Franck–Condon factors. An ultracold gas of He$_2$($a~^3\Sigma_u^+$) could enable a new generation of high-precision spectroscopic measurements with accuracies beyond the current state-of-the-art experimental limit.  

Highly accurate quantum chemical calculations for He$_2$($a~^3\Sigma_u^+$) predict a very deep potential well \cite{DawidHe2dimer} and support many vibrational levels. Relatively large rotational energy spacings \cite{VerdegayIJP2025,WirthPRA2025} in He$_2$($a~^3\Sigma_u^+$) may facilitate quantum-state-resolved scattering experiments. Nuclear spin-statistic theorem eliminates even rotational level of $^4$He$_2$($a~^3\Sigma_u^+$), reducing number of channels and makes He$_2$($a~^3\Sigma_u^+$) an ideal platform for benchmarking highly accurate quantum scattering calculations. 

Recent experimental studies have reported high precision spectroscopic measurements of He$_2$($a~^3\Sigma_u^+$) to provide benchmark data \cite{WirthPRA2025}. Experimental efforts have explored prospects for direct laser cooling of He$_2$($a~^3\Sigma_u^+$) \cite{VerdegayIJP2025}. Intermolecular interactions for He$_2$($a~^3\Sigma_u^+$)+He$_2$($a~^3\Sigma_u^+$) \cite{DibyenduHetetramer} and collision dynamics involving He$_2$($a~^3\Sigma_u^+$)+Ne/Ar \cite{DibyenduJCP2026} complexes have been investigated in both the cold and ultracold energy regimes. 

Extending these investigations to complexes involving a He$_2$($a~^3\Sigma_u^+$) and He(1~$^1\mathrm{S})$ represents a natural next step. Such systems offer an opportunity to investigate how metastable state influences few-body binding, molecular structure, and low-energy collision dynamics. In particular, whether the exceptionally weak interactions characteristic of helium can support bound states in a metastable helium trimer, and how its scattering properties compare with those of ground-state He$_3$, remain open questions.

In this letter, we address these questions by investigating the bound-state and low-energy scattering properties of He$_2$($a~^3\Sigma_u^+$)+He($^1\mathrm{S}$) complex in its lowest spin-polarized state, $a~^3\mathrm{A}'$. We benchmark potential energy surface (PES) using state-of-the-art single-reference coupled-cluster and multireference configuration-interaction methods, combined with extensive Gaussian basis sets. Independent electronic structure calculations consistently predict a linear equilibrium geometry for metastable He$_3$, characterized  by an extremely shallow potential well. Using this PES, we perform quantum close-coupling calculations, explicitly accounting for the fine structure of He$_2$($a~^3\Sigma_u^+$). We predict a single bound state in metastable He$_3$, with a binding energy of 9.423 mK. Remarkably, this state persists under substantial variations of the interaction potential. Our predictions provide a theoretical benchmark for the presence of a single quantum state in metastable He$_3$. Present results should motivate experimental investigation of vibrational binding energy and molecular structure potentially using advanced Coulomb-explosion imaging techniques \cite{KunitskiScience2015,VoigtsbergerNatCommn2014}.

\textit{Structural properties of metastable He$_3$}\textemdash We investigate the structure of metastable He$_3$ in its lowest spin-polarized $a~^3\mathrm{A}'$ electronic state. We compute a two dimensional (2D) PESs of He$_3$ trimer in Jacobi coordinates ($R, \theta$). Here $R$ is the distance from the center-of-mass of He$_2$($a~^3\Sigma_u^+$) to He($^1\mathrm{S}$), while $\theta$ is the angle between the diatomic molecular axis and the vector connecting the center-of-mass of dimer to He($^1\mathrm{S}$). Description of coordinates is given to the Supplemental Material (SM). We compute 2D PES under the rigid-rotor approximation for He$_2$($a~^3\Sigma_u^+$) and its bond length remains fixed at the equilibrium value, $r_{\mathrm{He_2}}=1.9894$ bohr \cite{DawidHe2dimer}.

To assess the accuracy of interaction potential and benchmark the PES, we compare results obtained in single reference coupled cluster calculations with single, double, and perturbative triple excitations CCSD(T) \cite{KnowlesJCP1993,KnowlesJCP2000}  with those obtained using the multireference configuration interaction method with single and double excitation (MRCI) \cite{KnowlesCPL1988,WernerJCP1988}, considering several basis sets. In addition to the available Hill basis sets \cite{VanMolPhys1999}, we employ a new basis set family \cite{dawid} especially formulated for describing the metastable state of helium. We calculate interaction energy of trimer with respect to He$_2$($a~^3\Sigma_u^+$)+He($^1\mathrm{S}$) asymptote. Details of the electronic structure calculations and basis sets are discussed in SM.

\begin{figure}[t]
  \includegraphics[width=\linewidth]{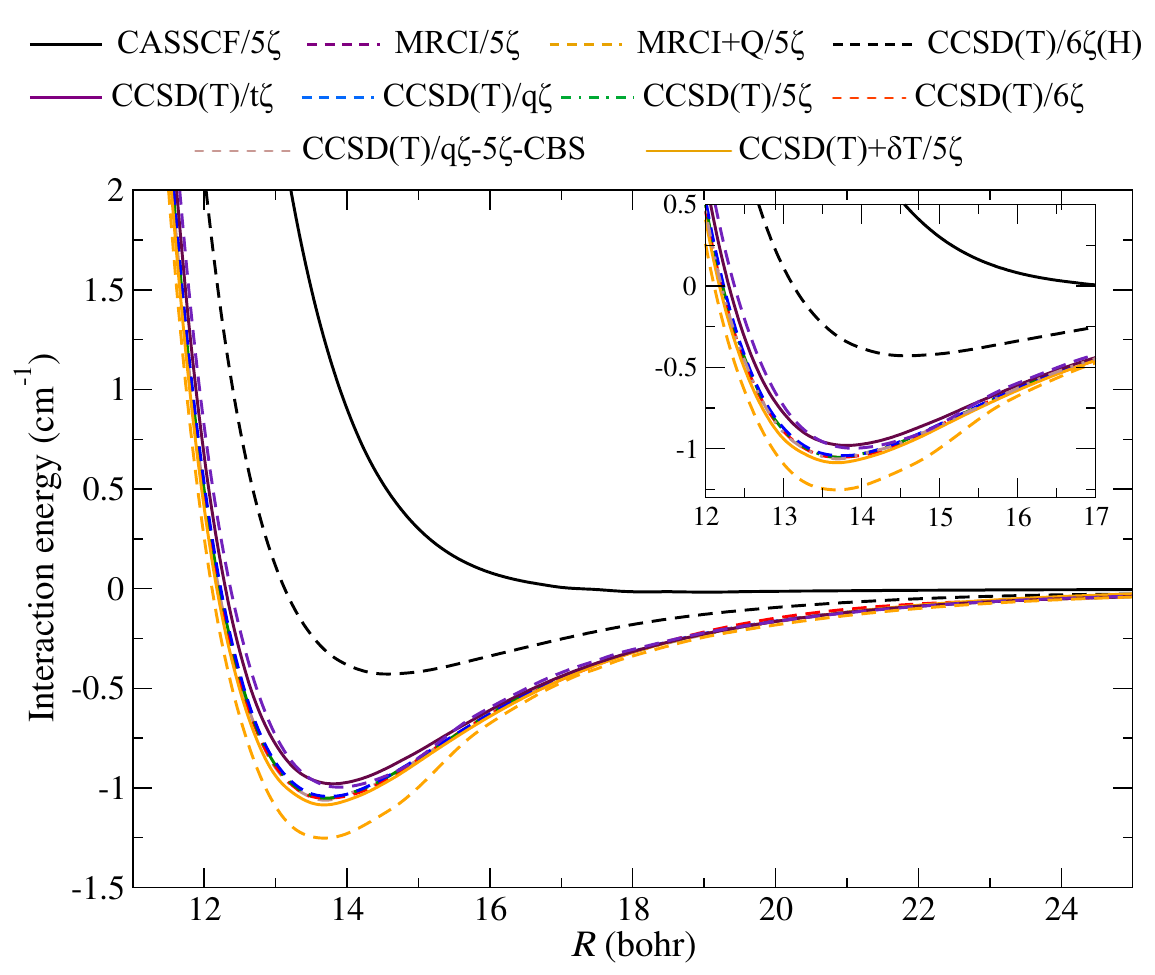}
  \caption{ Interaction energy as a function of Jacobi distance $R$ and $\theta=0^\circ$ for the lowest $a~^3\mathrm{A}'$ electronic state of He$_2$($a~^3\Sigma_u^+$)+He($^1\mathrm{S}$) van der Waals complex from non-relativistic Born-Oppenheimer calculations at different levels of theory. The inset represents the enlarged part around the equilibrium minimum.
  }\label{fig:1D}
\end{figure}

Our \textit{ab initio} calculations predict a linear equilibrium geometry of metastable He$_3$, with an extremely shallow potential well depth, approximately of $V_{\textrm{min}}=1.084$ cm$^{-1}$.
Figure \ref{fig:1D} presents 1D PES cuts for He$_2$($a~^3\Sigma_u^+$)+He($^1\mathrm{S}$) interactions in linear configuration. The CCSD(T) interaction energies obtained with new quadrupole-, quintuple-, and hexapole-$\zeta$ basis sets \cite{dawid} are in close agreement, with relative differences of less than 1\%. For triple-$\zeta$ basis, however, the relative difference increases to approximately 7\% . The CCSD(T) interaction energy obtained at the complete basis set (CBS) limit is also in close agreement with these results. In contrast, CCSD(T) calculation employing the hexapole-$\zeta$ Hill basis (H) \cite{VanMolPhys1999} set underestimates equilibrium well depth by approximately 50\%, indicating that this basis set provides an inadequate description of the interaction in the metastable He$_3$ system. Including a full triple excitation correction ($\delta$T) on CCSD changes the interaction energy at equilibrium by only 0.033 cm$^{-1}$, suggesting that contributions beyond perturbative triple excitations have a minor effect.

%These comparisons support the reliability of CCSD(T) method with the new basis sets for constructing the full 2D PES.

In addition, we investigate and assess the interaction energy for the linear geometry of metastable He$_3$ using multireference methods. Complete active-space self-consistent-field (CASSCF) calculations yield positive interaction energies in the vicinity of the equilibrium geometry, indicating that static electron correlation alone is insufficient to describe the binding in the He$_3$ trimer. In contrast, MRCI provides a reasonably accurate description of the potential well. Inclusion of the Davidson correction (+Q), the predicted well depth is $-1.251$ cm$^{-1}$, highlighting the importance of dynamical electron correlation. However, MRCISD is neither strictly size-consistent nor size-extensive, and the Davidson correction does not fully restore these properties. Thus, to provide an accurate description of the interaction, we employ CCSD(T) method to construct the 2D PES of the He$_3$ trimer. CCSD(T) is widely considered as the gold standard in \textit{ab initio} quantum chemistry calculations. We use aug-cc-pCV5Z basis set from a newly developed basis-set family specifically designed to describe the metastable state of He$_2$ dimer.

\begin{figure}[t]
  \includegraphics[width=\linewidth]{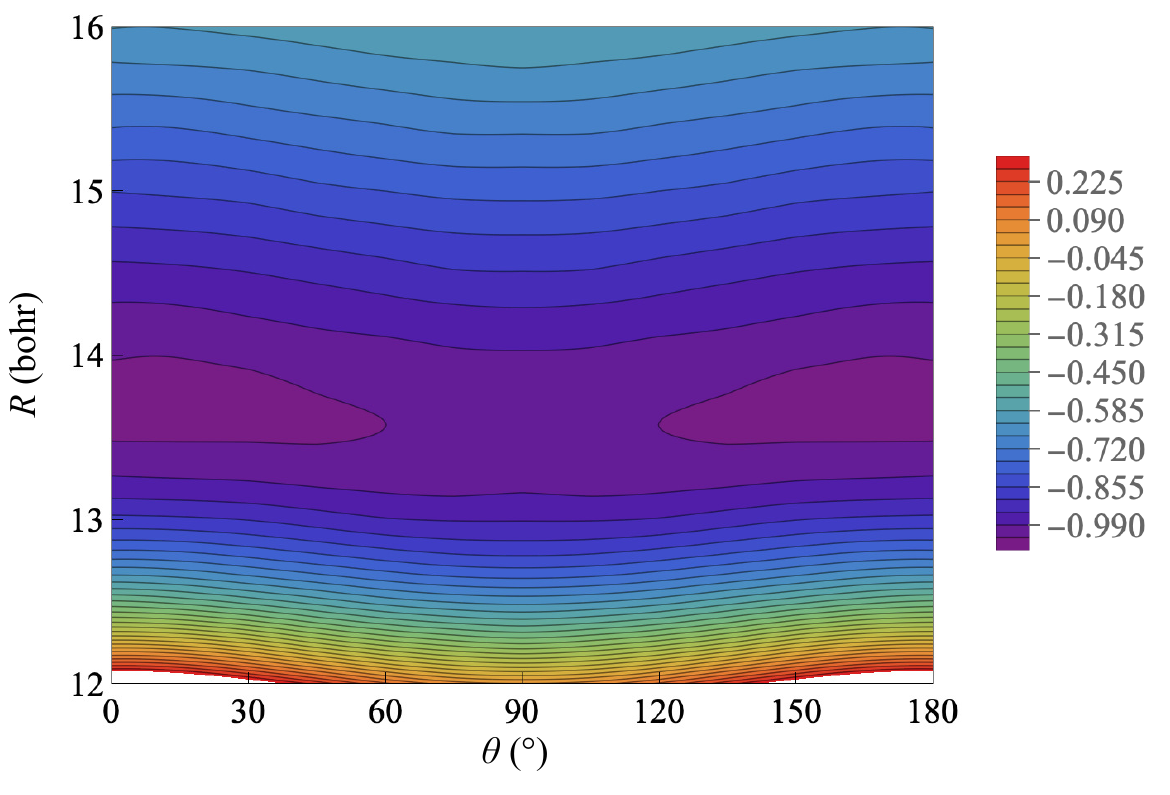}
  \caption{ A 2D contour PES for metastable He$_3$ trimer as a function of Jacobi coordinates $R$ and $\theta$. The colour scale indicates the interaction energy (in cm$^{-1}$) relative to the He$_2(a~^3\Sigma_u^+)$+He($^1\mathrm{S}$) dissociation threshold.
  }\label{fig:2D}
\end{figure}

In Figure \ref{fig:2D}, we present a 2D PES for the $a~^3\mathrm{A}'$ electronic state of He$_3$. The interaction potential is represented by the analytical expansion \cite{WernerJCP1989}, in which the PES is expanded in terms of Legendre polynomials. This representation is quite useful for the subsequent quantum close-coupling (CC) calculations. The analytic potential \cite{WernerJCP1989} reproduces the \textit{ab initio} interaction energies quite well. Across the entire 2D radial and angular grids, the mean difference between the analytic fit and the computed \textit{ab initio} interaction energies is less than 1\%. 
%2D PES for He$_2$($a~^3\Sigma_u^+$)+He($^1\mathrm{S}$) interaction is symmetric under interchange of two He atoms, corresponding to $\theta\rightarrow\pi-\theta$. Consequently, all odd-oder Legendre components vanish, and PES expansion contains only even order terms. 
For He$_3$ PES, the angular dependence of the interaction is relatively weak, so the isotropic component makes a substantial contribution compared to the anisotropic components.

We perform a full-dimensional geometry optimization of metastable He$_3$ in Jacobi coordinates at the CCSD(T) level. The optimized structure is linear, with $R=13.72$ bohr and $\theta=0^\circ$, a similar equilibrium property obtained from analytical PES. The potential well depth is lower than that of the ground-state He$_3$. This weak binding reflects the shallow dimer-atom interactions on the $a~^3\mathrm{A}'$ surface. Interaction strength of He$_2$($a~^3\Sigma_u^+$)+He($^1\mathrm{S}$) substantially weaker than the He$_2(a~^3\Sigma_u^+)$+Ne/Ar complexes \cite{DibyenduJCP2026}. Thus, metastable He$_3$ exhibits an unusually shallow potential well even compared with other van der Waals complexes involving He$_2(a~^3\Sigma_u^+)$. This makes it a distinctive system for investigating bound states and low-energy collision dynamics.

\textit{Bound states and scattering properties}\textemdash
We carry out quantum coupled-channel calculations on the $a~^3\mathrm{A}'$ surface to compute bound state and scattering properties of metastable He$_3$. In coupled-channel calculations, we consider only $^4$He isotopes. The molecule-atom Hamiltonian in space-fixed coordinates is given by 
\begin{equation}
    H=-\frac{\hbar^2}{2\mu R}\frac{\partial^2}{\partial R^2}R+ \frac{L^2}{2\mu R^2} + V(R, \theta) + H_{\mathrm{He}_2}, 
    \label{eq:totH} 
\end{equation}
where the first term is relative kinetic energy of molecule+atom collision complex, and $\mu$ is the reduced mass. The second term is the centrifugal term associated with the end-over-end rotational angular momentum $\boldsymbol{L}$ of the interacting pair. The molecular rotational Hamiltonian and constant parameters for He$_2$($a~^3\Sigma_u^+$) are given in SM. We solve coupled-channel equations subject to bound state and scattering boundary conditions using the \textsc{Bound} \cite{HutsonCompPhy1994} and \textsc{Molscat} \cite{HutsonCompPhy2019} packages, respectively. Details of the propagation parameters for both packages are provided in SM.   

For He$_2$ in $a~^3\Sigma_u^+$ electronic state, each rotational level with quantum number $N$ is split into three fine structure components due to spin-spin and spin-rotation interactions. These fine-structure levels are explicitly included in the coupled-channel calculations. In pure Hund's case (b) limit, fine structure levels of He$_2$( $a~^3\Sigma_u^+$) follow the energy ordering $F_2<F_1<F_3$ \cite{VerdegayIJP2025,WirthPRA2025} for all values of $N$. In this work, we include rotational levels of He$_2$($a~^3\Sigma_u^+$) up to $N\le 9$ and solve Eq \ref{eq:totH} with coupled-channel basis sets described in Ref. \cite{DibyenduJCP2026}. 

\begin{table}[t]
\caption{\label{tab:bound} Bound-state energies ($E_b$ in mK) of metastable He$_3$, including the fine structure of He$_2$, for different potential scaling factors $\gamma$.}
\begin{ruledtabular}
\begin{tabular}{ccc}
% &\multicolumn{2}{c}{global geometry}&\multicolumn{2}{c}{local geometry}\\
   $\gamma$ & $L$ & $E_b$ (mK)   \\
\hline
  0.95& 0 & -2.411 \\ 
  0.98& 0 & -6.609  \\
  1.00 & 0 & -9.423  \\ 
  1.03 & 0 & -13.649 \\
  1.05& 0 & -16.446  \\ 
 \end{tabular}
\end{ruledtabular}
\end{table}

Our coupled-channel calculations reveal a single bound state of the metastable He$_3$ trimer. This state is found for a total angular momentum $J=1$ and has end-over-end orbital angular momentum $L=0$, and positive parity. Its binding energy is 9.423 mK, measured relative to the $F_2$ fine-structure threshold of He$_2$($a~^3\Sigma_u^+$). The corresponding eigenvector is entirely dominated by a single channel with $L=0$. No bound states are found in the $J=0$ or higher-$J$ sectors within the basis and parameter ranges considered.

To assess the sensitivity of the bound state to the interaction potential, we scale the potential by slightly varying the parameter $\gamma$ in the ranges from $0.95$ to $1.05$. The resulting bound-state energies are summarized in Table~\ref{tab:bound}. We scale the potential according to $V(R,\theta)\rightarrow \gamma V(R,\theta)$. We find that a single bound state persists throughout the range of scaling parameters considered, although its binding energy changes with the strength of the interaction. This persistence indicates that the existence of the bound state is robust against moderate variations in the potential, rather than being an artifact of a particular choice of potential parameters. To validate the methodology used in the present work, we employ the same level of theory (both for the \textit{ab initio} calculations and for the bound states) for studying the ground state of He$_3$. We find the two experimentally observed bound states for the ground states of helium trimer.  Further details are provided in SM. 

%The ground state He$_3$ is deeply bound and exhibit strong anisotropic behavior compared to that of the excited metastable He$_3$.
%Using a combination of highly accurate electronic structure calculations followed by quantum close-coupling methods,
%The persistence of the metastable He$_3$ bound state under potential scaling, together with the agreement with reported results for ground-state He$_3$, supports the reliability of our predictions.

Metastable He$_3$ trimer supports a bound state, as does the ground state of He$_2$ dimer \cite{LuoJCP1992}. However, ground state of He$_3$ trimer supports two bound state \cite{HiyamaPRA2012}: a deeply bound ground state and a weakly bound excited state with Efimov character \cite{KunitskiScience2015}. Our computational approach yields two bound states of the ground-state He$_3$ trimer, as described in detail in the SM.  In contrast,  we identify only one bound state of metastable He$_3$.

%within the angular-momentum sectors and basis considered. 

%Although the binding energy of this state is larger than that of the excited Efimov state of ground-state $^4$He$_3$, it is substantially smaller than the binding energy of the ground state. 

%Whether this bound state of metastable He$_3$ exhibits Efimov character remains an open question and is beyond the scope of the present study. xperimentally measured binding energies are 131.84 mK and 2.6505 mK, respectively \.

Notably, the binding energy of metastable He$_3$ is much smaller than the potential well depth, \(E_b \ll V_{\mathrm{min}}\). In particular, the binding energy accounts for only approximately \(0.6\%\) of the well depth. This indicates that a weakly bound trimer state may lie energetically close to the dimer-atom dissociation threshold. Moreover, $E_b$ is substantially smaller than the rotational constant, \(b=7.589\ \mathrm{cm}^{-1}\) \cite{VerdegayIJP2025}, of He$_2$($a~^3\Sigma_u^+$), with \(E_b/b \approx 8.6\times10^{-4}\), highlighting the pronounced separation between the trimer binding and molecular rotational energy scales. However, these energy-scale comparisons alone do not establish universal behavior, which is independent on the specific shape or microscopic details of the short-range interaction. Our preliminary coupled-channel calculations for He$_2$($a~^3\Sigma_u^+$)+He($^1\mathrm{S}$) collision yield an dimer-atom scattering length of $a_s=7.495~a_0$, which  is smaller than the corresponding van der Waals length, $r_{\mathrm{vdW}}=12.043~a_0$. Thus, $|a_s|$ is not large compared to the characteristic van der Waals length,  and the usual large-scattering-length criterion underlying the zero-range universal description \cite{EfimovNuPhy1973energy,NaidonRPP2017} is not fulfilled. Nevertheless, further investigation is needed to assess the universality of the bound state, which is beyond the scope of the present study.

 %an experimental determination of scattering length would clarify universality of bound state and benchmark predicted $a_s$ for He$_2$($a~^3\Sigma_u^+$)+He($^1\mathrm{S}$) collision. 

\begin{figure}[t]
  \includegraphics[width=\linewidth]{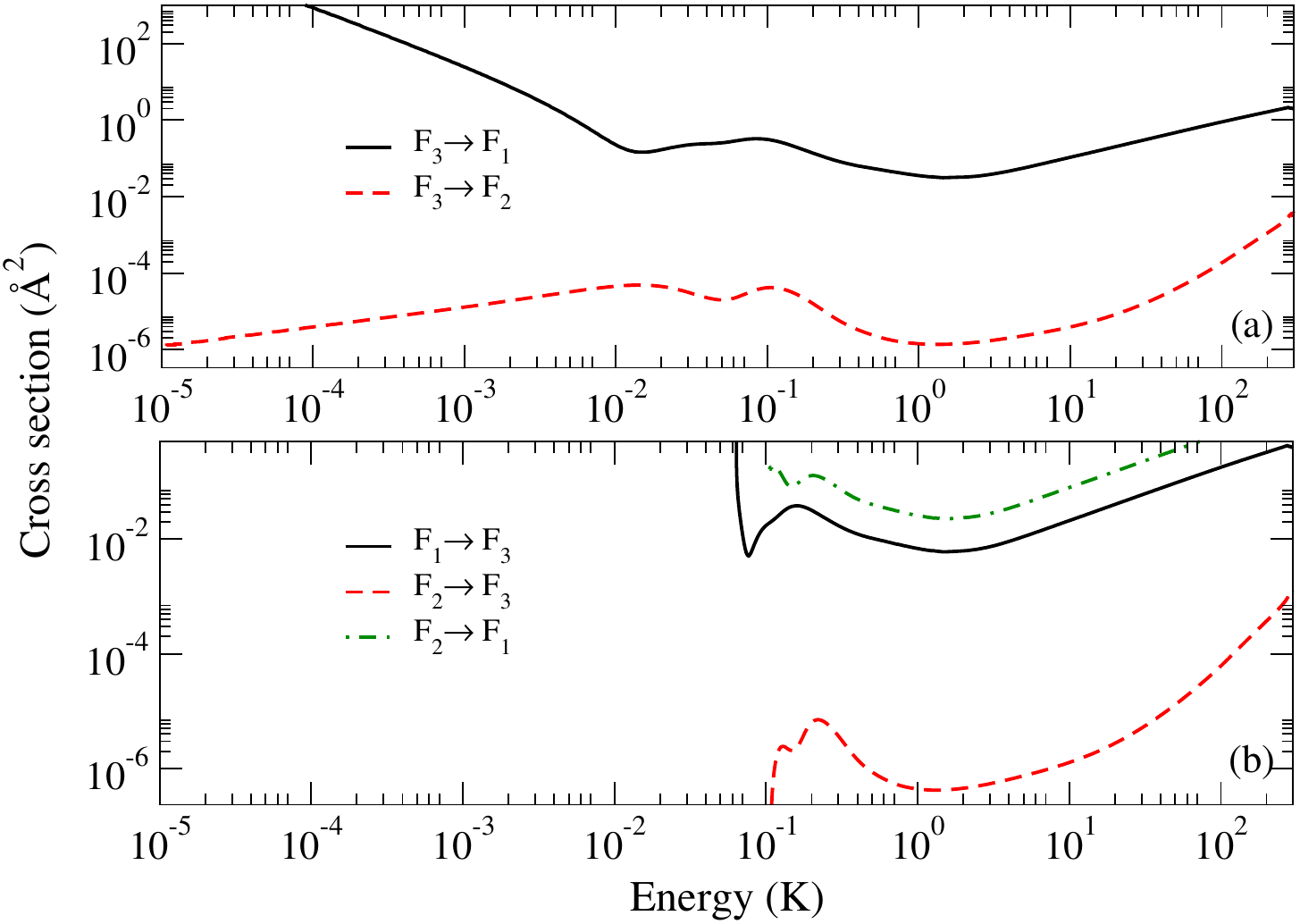}
  \caption{ He$_2$+He scattering cross sections (in $\mathrm{\AA}^2$) for quenching or de-excitation and excitation transitions within $N=1$ are shown in panel (a) and (b), respectively. Both cross-section and collision energy axes are on a logarithmic scale. 
  }\label{fig:cross}
\end{figure}

%The bound-state spectrum can provide important insight into He$_2$($a~^3\Sigma_u^+$)+He($^1\mathrm{S}$) collision dynamics.
In Figure \ref{fig:cross}, we present inelastic scattering cross sections for quenching and excitation transitions of He$_2$($a~^3\Sigma_u^+$)+He($^1\mathrm{S}$) collisions within $N=1$ rotational level for He$_2$. The scattering cross sections are converged with respect to total angular momentum up to $J=19$ at a collision energy of 1 K. At ultralow collision energies, quenching $F_3\rightarrow F_1$ transition has a substantial cross section,  while cross section for $F_3\rightarrow F_2$ transition is strongly suppressed and approaches to zero. In ultracold energy limit, He$_2$($a~^3\Sigma_u^+$)+He($^1\mathrm{S}$) collision is dominated by $s$-wave, and cross section exhibits threshold behavior governed by the Wigner threshold laws \cite{krems2018molecules}. However, pronounced suppression in cross section for $F_3\rightarrow F_2$ transition may reflect forbidden or indirectly coupled transition. Such transition is observed in C+He \cite{StaemmlerJPB1991} and O$_2$+H$_2$ \cite{SimonScience2013} collisions. At higher collision energies, between 1 and 300 K,  both quenching and excitation cross sections exhibit a similar energy dependences.

The inelastic cross sections for He$_2$($a~^3\Sigma_u^+$)+He($^1\mathrm{S}$) collisions exhibit a relatively smoothly dependence on collision energy without a dense spectrum of resonances. This behavior is consistent with the sparse bound-state spectrum of metastable He$_3$. Nevertheless, a resonance feature appears in cross sections near 0.1 K. In comparison, the closely related system He$_2$($a~^3\Sigma_u^+$)+Ne($^1\mathrm{S}$) \cite{DibyenduJCP2026} has richer resonance structure in scattering cross sections due to the different interaction potentials and associated quasibound-state spectra of the two systems.  To elucidate the origin of resonance, we perform a partial wave analysis \cite{PlompJPCL2024}. 

\begin{figure}[t]
  \includegraphics[width=\linewidth]{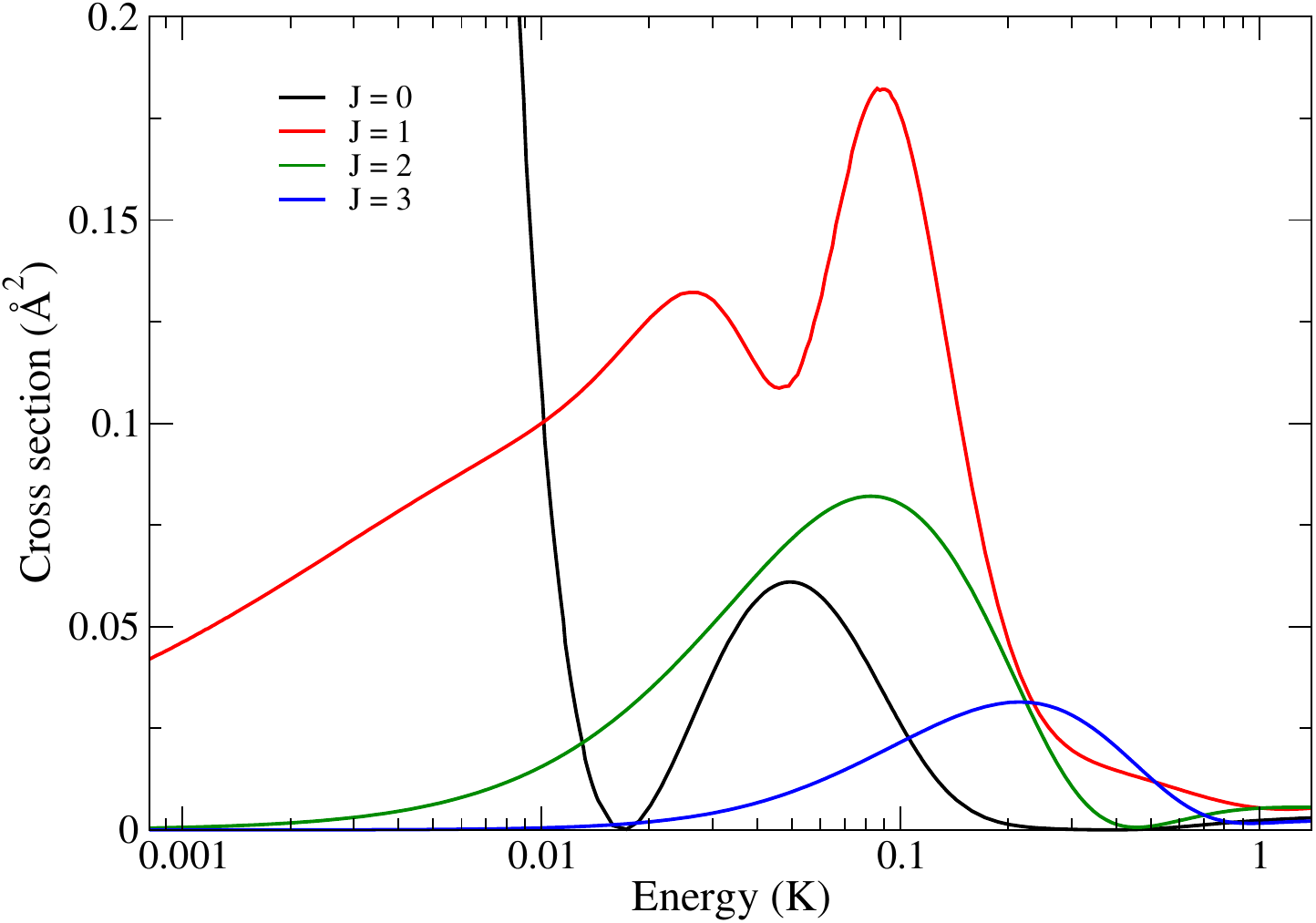}
  \caption{ Inelastic scattering cross sections as a function of energy in $F_3\rightarrow F_1$ quenching transition for the lowest four values of $J$. 
  }\label{fig:res}
\end{figure}

In Figure \ref{fig:res}, we plot cross sections of $F_3\rightarrow F_1$ transition corresponding to $J=0-3$. Rapid variation in cross section for individual $J$ with distinct valleys and peaks may correspond to a scattering resonance. We find $J=1$ state has a predominant contribution near collision energy 0.1 K and attribute the primary source of resonance feature in Figure \ref{fig:cross}. We identify two quasi-bound states for $J=1$ with binding energies close to 0.206 K, when internal energy of initial $F_3$ level, $E_{F_3}=0.105$ K is included. These quasi bound states may be temporarily trapped behind centrifugal barriers in the relevant $J=1$ channels, giving rise to the observed resonant structure.

%which may arise from presence of a quasi-bound states supported by the attractive He$_2$($a~^3\Sigma_u^+$)+He($^1\mathrm{S}$) interaction potential in combination with the centrifugal barrier. 

Metastable helium trimer is a paradigmatic molecular system.  By using state-of-the-art \textit{ab initio} quantum chemical methods, we predict a linear equilibrium geometrical shape of trimer. It is extremely weekly bound with a potential well depth of approximately 1.084 cm$^{-1}$. This could be the first most weakly bound three-body van der Waals molecule. Quantum close-coupling calculations that explicitly accounts for fine structure of He$_2$($^3\Sigma_u^+$) predicts a single bound state in metastable helium trimer. The bound state is obtained only for total angular momentum $J=1$ of collision complex, and $L=0$ without admixture of higher partial waves. We checked and confirmed that no additional bound state obtained in $J=0$ and higher $J$-sectors within the basis and parameter ranges considered. Inelastic scattering cross sections exhibit a smooth variation of energy dependences, with no discernible resonance enhancement. This behavior could be associated with a sparse bound-state spectrum of trimer. 

%Our quantum close-coupling calculations predict $s$-wave scattering length $7.501~a_0$ is eventually lower than that of the computed van der Waals length $12.043~a_0$. It suggests that the bound state may not be present in the universal regime \cite{EfimovNuPhy1973energy}.  

Metastable He$_3$ is a promising system, even if our knowledge is based exclusively on theory. An experimental investigation is the next natural step in determining chemical structure and bound state properties of He$_3$. Exceptionally weak binding and the presence of only one predicted bound level pose challenges for conventional bound–bound rotational and vibrational spectroscopy for this extremely fragile He$_3$ cluster. Matter-wave diffraction \cite{SchollkopfScience1994}, which has been used to investigate spatial structure of the ground-state He$_3$ \cite{BruhlPRL2005}, offers a possible route to probing the geometry of metastable He$_3$. Advanced state-of-the-art Coulomb explosion imaging technique may be a complementary approach \cite{KunitskiScience2015,VoigtsbergerNatCommn2014}. We anticipate that this technique could be useful to detect, and resolve the structure and bound state properties of metastable He$_3$ complex. Our calculations provide an initial testbed and guide future experimental investigations.

%However, traditional experimental approaches such as vibrational and rotational spectroscopy may not be applied to  trimer having very low binding energy. Some geometrical features of metastable He$_3$ can be measured by matter wave diffraction, which is successful for the ground state of He$_3$. However, a complete description for structure and bound state features can be obtained by advanced state-of-the-art Coulomb explosion imaging. 

%The metastable He$_3$ trimer remains experimentally unexplored, and its predicted structure and bound-state properties await direct verification. .  Coulomb-explosion imaging could provide a complementary approach by probing the nuclear geometry through correlated fragment momenta following rapid ionization. Whether this technique can resolve the structure of such a weakly bound metastable complex remains an open experimental question. The present predictions provide specific structural, spectroscopic, and scattering observables against which future measurements of metastable He$_3$ can be tested. 

%---------------------------------------------------------------
%\section{Supplementary material}
%See the supplementary material for explicit matrix elements of spin-rotations and spin-spin operators in Eq \ref{eq:Hmol} to facilitate reproducibility of the calculations.

%---------------------------------------------------------------------------------------------

\begin{acknowledgments}
We are thankful to Andrea Simoni for valuable discussions.
We gratefully acknowledge support of the European Union through the Marie Sk{\l}odowska-Curie Action (MSCA)-Postdoctoral Fellowship HighSpin$\_$Helium, under Grant Agreement No. 101271813. We acknowledge \textsc{SCAPHYR} for computational facilities at Université de Rennes, France.
\end{acknowledgments}

\begin{dataavailability}
The fitting coefficients for He$_2$+He PESs, and the corresponding cross sections data (Figure \ref{fig:cross}) are openly available in ZENODO Repository at doi link: https://doi.org/10.5281/zenodo.24753135. All the \textit{ab initio} data for interaction potentials are available from the authors upon reasonable request.
\end{dataavailability}

\bibliography{reference.bib}
%\end{multicols}
%\input{SI.tex}
\end{document}

%\title{Ultracold CaF and Ca interactions in ground and excited electronic states: implications to chemical reactions}
\title{Supplemental Material: He$_2(a~^3\Sigma_u^+)$+He complex: A promising system for cold and ultracold collision studies}

\author{Dibyendu Sardar}
\email{dibyendu.sardar@univ-rennes.fr}
\affiliation{Institut de Physique de Rennes, CNRS-Universit\'e de Rennes, UMR 6251, F-35000 Rennes, France}
%\affiliation{JILA, University of Colorado, Boulder, Colorado 80309, USA}

%\altaffiliation{Present address: Faculty of Physics, University of Warsaw, Poland}
\author{Fran\c{c}ois Lique}
\email{francois.lique@univ-rennes.fr}
\affiliation{Institut de Physique de Rennes, CNRS-Universit\'e de Rennes, UMR 6251, F-35000 Rennes, France}

\date{\today}% It is always \today, today,

\maketitle
\begin{figure}[t]
  \includegraphics[width=0.85\linewidth]{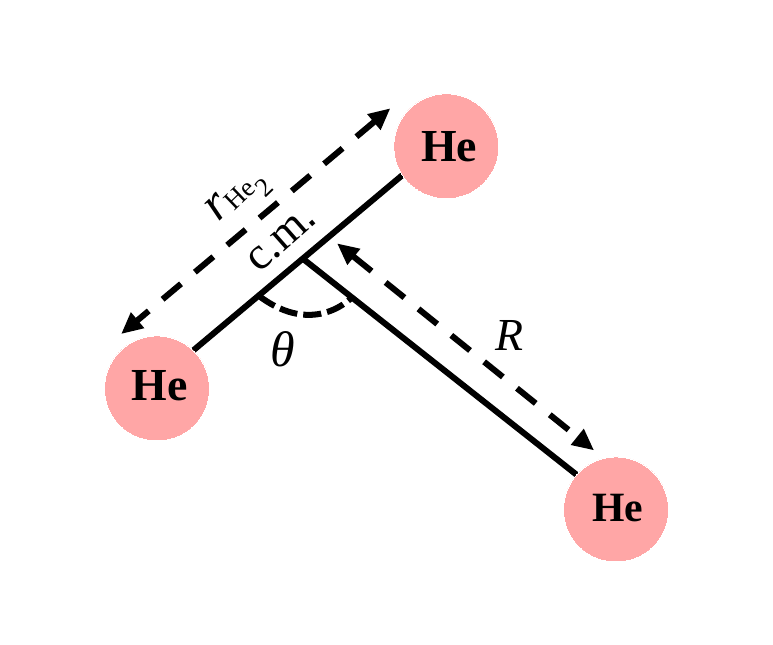}
  \caption{ A schematic diagram for the Jacobi coordinates of He$_3$ trimer.
  }\label{fig:schematic}
\end{figure}

%\section{Computational methodology}
\textit{Electronic structure calculations}\textemdash
We compute PES for the lowest spin-polarized $a~^3\mathrm{A}'$ electronic state of He$_2$($a~^3\Sigma_u^+$)+He($^1\mathrm{S}$) van der Waals complex using state-of-the-art quantum-chemical methods. Electronic structure calculations are performed using the 2015 \textsc{Molpro} software package \cite{WernerJCP2020}, under C$_\mathrm{s}$ point group symmetry. We present a triatomic He$_2$($a~^3\Sigma_u^+$)+He($^1\mathrm{S}$) complex in Jacobi coordinates. In Figure \ref{fig:schematic}, we show a schematic diagram for Jacobi coordinates,  where $R$ is the distance between the centre-of-mass (c.m.) of He$_2$ and He, and $\theta$ is the angle between the molecular axis and c.m. to He which defines rotation of He$_2$ dimer. We compute PES for He$_2$($a~^3\Sigma_u^+$)+He($^1\mathrm{S}$) complex under rigid-rotor approximation for He$_2$($a~^3\Sigma_u^+$) dimer and its bond length remains fixed at the equilibrium value, $r = 1.9894$ bohr \cite{DawidHe2dimer}.

In the electronic structure calculations, we use polarized, double augmented, correlation-consistent basis sets. In particular, we consider aug-cc-pCVXZ (where X = T, Q, 5, 6) \cite{dawid}, hereafter, we represent acpCVXZ. These new class of basis sets were developed to accurately describe He$_2$($a~^3\Sigma_u^+$) when one He atom is in the ground state and another in the metastable $2~^3\mathrm{S}$ state. For a comparison of the interaction energy, we consider correlation-consistent polarized valence sextuple-$\zeta$ Hill basis set \cite{VanMolPhys1999} for He with diffuse augmenting function. The Hill basis is effective to describe for the ground state interaction in He.

To describe equilibrium geometry of He$_3$ trimer in its lowest $a~^3\mathrm{A}'$ electronic state, we consider single-reference CCSD(T), and multireference MRCI methods. The interaction energies are obtained using the super-molecule approach with the basis set superposition error (BSSE) corrected by using the Boys-Bernardi counterpoise correction \cite{BoysMolPhys1970}. For the equilibrium configuration of trimer, we consider leading electron-correlation correction to the interaction energy. This correction accounts for the contribution of iterative full triple excitations in the coupled-cluster method and is given as
\begin{equation}
    \delta V_{\mathrm{CCSDT}}^{\mathrm{acpCVTZ}} =  V_{\mathrm{CCSDT}}^{\mathrm{acpCVTZ}} -  V_{\mathrm{CCSD(T)}}^{\mathrm{acpCVTZ}}.
\end{equation}
Here, $V_{\mathrm{CCSDT}}^{\mathrm{acpCVTZ}} $ and $V_{\mathrm{CCSD(T)}}^{\mathrm{acpCVTZ}}$ present the interaction energy computed using the coupled-cluster with single, double, and triple excitations (CCSDT) and  CCSD(T) methods. In both methods, we consider acpCVTZ basis set. Final interaction energies, considering $\delta V_{\mathrm{CCSDT}}^{\mathrm{acpCVTZ}}$ correction, are expressed as
\begin{equation}
    V = V_{\mathrm{CCSD(T)}}^{\mathrm{acpCV5Z}}+\delta V_{\mathrm{CCSDT}}^{\mathrm{acpCVTZ}}
\end{equation}
However, for computing full 2D PES of trimer, $V(R,\theta)$, we employ CCSD(T) method with acpCV5Z basis. 

To construct 2D PES, we consider dense grids of $R$ and $\theta$. We consider 50 values of $R$, spanning from 6-40 bohr. For the angular coordinate, we consider 13 grid points, covering range $(0^\circ \leq \theta \leq 90^\circ)$ in increments of $7.5^\circ$. Since He$_2$ is homonuclear, the remaining angular range is related to this interval by symmetry, and therefore no additional \textit{ab initio} calculations are required. We first generate converged partially open-shell restricted Hartree–Fock (RHF) orbitals on the 2D $(R,\theta)$ grid. These orbitals serve as reference orbitals for subsequent CCSD(T) calculations. To ensure convergence, we perform the \textit{ab initio} calculations using a sequential distance-loop procedure, starting from well-behaved RHF orbitals at large intermolecular separations. The convergent wavefunction at each distance is then used as the initial guess for the CCSD(T) calculation at the next shorter distance.

We express 2D interaction potential in terms of Legendre polynomials $P_\lambda(^.)$: 
\begin{equation}
    V(R,\theta) = \sum_{\lambda=0}^{\lambda_{\text{max}}} V_\lambda (R) P_\lambda(\cos\theta), \label{eq:2Dpes}
\end{equation}
where, $\lambda$ denotes the order of the polynomials, and $V_\lambda(R)$ represents the corresponding Legendre expansion coefficient of the 2D PES. The Legendre expansion is useful for the subsequent quantum close-coupling calculations. To ensure a correct behaviour of PES expansion at long-range separations, the radial expansion coefficients are fitted to the following functional form \cite{DibyenduJCP2026}: 
\begin{equation}
\begin{aligned}
  V_\lambda (R)=& e^{-a_1^\lambda R}(a_2^\lambda + a_3^\lambda R + a_4^\lambda R^2 +a_5^\lambda R^3) \\
  &-\frac{1}{2}\left[1+\tanh\left(\frac{R}{R_{\mathrm{ref}}}\right)\right]\left( \frac{C_6^\lambda}{R^6} + \frac{C_8^\lambda}{R^8} + \frac{C_{10}^\lambda}{R^{10}}\right).
  \end{aligned}
\end{equation}
Here, $a_i^\lambda$ denote the short-range fitting parameters, while $C_n^\lambda$ are the coefficients of inverse-power terms $R^{-n}$. The hyperbolic tangent function ensures a smooth transition between the short-range region ($0<R<R_{\text{ref}}$), where \textit{ab initio} energies are computed, and long-range region ($R>R_{\text{ref}}$), where the PES is extrapolated.

%We follow the same procedure to compute full 2D PES for the ground state helium trimer, obtained by the interaction of He$_2 (X~^1\Sigma^+_g)$+He($^1\mathrm{S}$).

%In the present work, we use PES averaged over ground vibrational state $v=0$ for He$_2$ in He$_2$+Ne collision dynamics.
\textit{Quantum close-coupling approach}\textemdash The Hamiltonian for the interacting He$_2$($a~^3\Sigma_u^+$)+He($^1\mathrm{S}$) in the space-fixed (SF) coordinates is given as
\begin{equation}
    H=-\frac{\hbar^2}{2\mu R}\frac{\partial^2}{\partial R^2}R+ \frac{L^2}{2\mu R^2} + V(R, \theta) + H_{\mathrm{He}_2}. 
    \label{eq:totH} 
\end{equation}
Here, the first term describes the relative kinetic energy of molecule+atom collision complex, while $\frac{L^2}{2\mu R^2}$ is the centrifugal term that describes the end-over-end rotation. $\boldsymbol{L}$ is the end-over-end rotational angular momentum and $\mu$ is the reduced mass mass of collision complex. Molecular rotational Hamiltonian for He$_2$($a~^3\Sigma_u^+$) is expressed as
\begin{equation}
    H_{\mathrm{He}_2}= bN^2  + \frac{2}{3}\lambda_0(3S^2_\xi-S^2) + \gamma_0\boldsymbol{N.S} ,
    \label{eq:Hmol}
\end{equation}
here, \(b=7.589~\mathrm{cm}^{-1}\) is the rotational constant of He\(_2\) in its \(a~^3\Sigma_u^+\) electronic state \cite{VerdegayIJP2025}. The constants \(\lambda_0=-0.036~\mathrm{cm}^{-1}\) and \(\gamma_0=-0.0000807~\mathrm{cm}^{-1}\) represent the spin-spin and spin-rotation coupling strengths, respectively, for this state \cite{VerdegayIJP2025}. The operator \(S_\xi\) denotes the projection of the electronic spin \(\mathbf{S}\) onto the He\(_2\) molecular axis.

We calculate bound states of He$_2$($a~^3\Sigma_u^+$)+He($^1\mathrm{S}$) complex by solving coupled-channel equations using \textsc{Bound} package \cite{HutsonCompPhy1994} with a coupled-channel basis. The coupled-channel basis set is described in Ref. \cite{DibyenduJCP2026}. Coupled-channel  bound-state spectrum of He$_2$($a~^3\Sigma_u^+$)+He($^1\mathrm{S}$) complex is obtained over the energy range $ E_{\mathrm{min}}=-1.5~\mathrm{cm}^{-1}$ to $E_{\mathrm{max}}=0.5~\mathrm{cm}^{-1}$. 

Scattering calculations are performed using the \textsc{Molscat} package \cite{HutsonCompPhy2019}. Such calculations provide the scattering \textit{\textbf{S}} matrix for a single value of the collision energy. From the \textit{\textbf{S}} matrix, we compute state-to-state scattering cross sections with collision energies ranging from $10~ \mu$K-350 K. To ensure convergence in cross sections, we include rotational basis of He$_2$  up to $N=9$. The maximum total angular momentum quantum number is increased until convergence is achieved within the DTOL and OTOL tolerances. We compute $s$-wave scattering length $a_s$ from the diagonal element of \textit{\textbf{S}} in the incoming channel $S_{00}$ \cite{HutsonNJP2007}
\begin{equation}
    a_s(k_0)=\frac{1}{ik_0}\left(\frac{1-S_{00}(k_0)}{1+S_{00}(k_0)}\right).
\end{equation}
Here, $k_0$ is the incoming wave number and is related to the collision energy as $E={\hbar^2k_0^2}/({2\mu})$. Scattering length becomes constant at very low collision energy. In the present work, scattering length calculations are performed for $E/k_b=10$ nK, where $k_b$ is the Boltzmann constant. 

For both bound-state and scattering calculations, we employ the Manolopoulos diabatic modified log-derivative propagator \cite{ManolopoulosJCP1986}. In both cases, we start propagation at $R_{\mathrm{min}}=6$ bohr. For bound-state calculations, the matching radius is set to $R_{\mathrm{mid}}=10.5$ bohr, with an outer propagation limit of $R_{\mathrm{max}}=80$ bohr. We set a step size of 0.001 bohr. For scattering calculations, the outer propagation limit is set to $R_{\mathrm{max}}=7500$ bohr in the ultracold and cold regimes and reduced to 200 bohr at higher collision energies. 

\textit{Properties of ground state He$_3$}\textemdash To validate the methodology employed, we also investigate the ground state He$_3$ trimer formed by the interactions of He\(_2\)(\(X~^1\Sigma_g^+\))+He(\(^1\mathrm{S}\)). We compare results for the ground-state He$_3$ trimer with previous calculations. This approach provides a useful benchmark to establish and validate our theoretical computational methodologies before extending to the metastable system. This additional analysis tests the accuracy of the coupled-cluster based interaction potential and the reliability of the coupled-channel method in describing three-body interactions and bound states for He$_2$($a~^3\Sigma_u^+$)+He($^1\mathrm{S}$) van der Waals complex. 

\begin{figure}[b]
  \includegraphics[width=\linewidth]{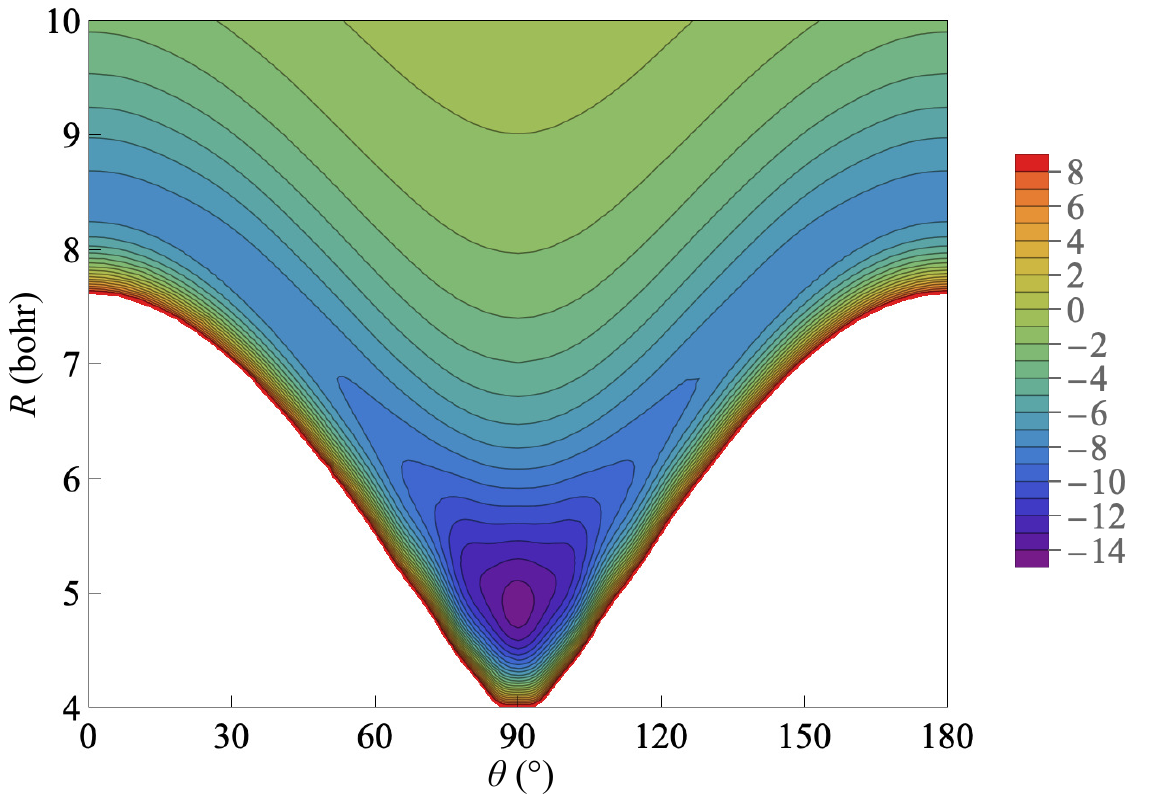}
  \caption{A 2D contour PES for the ground state of He$_3$ trimer in Jacobi coordinates. The colorbar presents interaction energy in cm$^{-1}$.
  }\label{fig:2Dground}
\end{figure}

We compute 2D PES for the ground state He$_3$ trimer using the CCSD(T) method with d-aug-cc-pV5Z basis for He in Jacobi coordinates. We use rigid rotor approximation for He$_2$($X~^1\Sigma_g^+$) and its bond length remains fixed at the equilibrium value \cite{CencekPRL2005}. In Figure \ref{fig:2Dground}, we present a 2D contour PES for the ground $X~^1\mathrm{A}'$ electronic state of He$_3$. The PES exhibits strong angular dependence, indicating a highly anisotropic character. The global equilibrium minimum for He$_3$ occurs at a T-shaped geometry, with $\theta=90^\circ$, and $R=4.880$ bohr. We predict a potential well depth of 14.563 cm$^{-1}$ with respect to He$_2$($X~^1\Sigma_g^+$)+He($^1\mathrm{S}$) asymptote. Similar equilibrium geometry is obtained in full-dimensional geometry optimization at the CCSD(T) level. The equilibrium geometry obtained within our approach corresponds to a nearly equilateral He$_3$ configuration. Specifically, the Jacobi coordinates $r=5.600$ bohr, $R=4.880$ bohr, and $\theta=90^\circ$ yield interatomic distances of 5.600, 5.626, and 5.626 bohr. These values agree closely with equilateral equilibrium geometry reported in Ref. \cite{Albernaz2016}. For comparison, a well depth of 17.222 cm$^{-1}$ was reported at CCSD(T)/d-aug-cc-pVQZ \cite{Albernaz2016} with respect to He$_2$($X~^1\Sigma_g^+$)+He($^1\mathrm{S}$) asymptote, including a small nonadditive three-body contribution 0.581 cm$^{-1}$. This value is larger than our calculated well depth. The discrepancy may partly arise from differences in basis-set quality and the treatment of basis-set superposition error. In contrast to the d-aug-cc-pVQZ basis set used in Ref. \cite{Albernaz2016}, our calculations employ the larger d-aug-cc-pV5Z basis set and explicitly account for BSSE through counterpoise correction. 

We use the calculated 2D PES to determine the bound states of the ground-state He$_3$ complex by solving the coupled-channel equations with the \textsc{Bound} package \cite{HutsonCompPhy1994}. We find one bound state on the unscaled PES, whereas a slight scaling of the PES supports a second bound state. Bound states energies are referenced with respect to He$_2$($X~^1\Sigma_g^+$)+He($^1\mathrm{S}$) asymptote. Although the calculated binding energies differ from previously reported values, our methods predict the existence of two bound states. Qualitative agreement supports the use of our approach to investigate the lowest spin-polarized $a~^3\mathrm{A}'$ electronic state of He$_3$ trimer.

%-------------------------------------------------------------------

\bibliography{reference.bib}